\documentclass[aip,reprint]{revtex4-2}
\usepackage{graphicx}
\usepackage{upgreek} 
\usepackage{wasysym} 

\begin{document}

\title{Planar scanning probe microscopy enables vector magnetic field imaging at the nanoscale}

\author{Paul Weinbrenner}
\affiliation{University of Rostock, Institute for Physics, 18059 Rostock, Germany}
\affiliation{University of Rostock, Department of Life, Light and Matter, 18059 Rostock, Germany}

\author{Patricia Klar}
\affiliation{Fraunhofer Institute for Applied Solid State Physics, 79108 Freiburg, Germany}

\author{Christian Giese}
\affiliation{Fraunhofer Institute for Applied Solid State Physics, 79108 Freiburg, Germany}

\author{Luis Flacke}
\affiliation{Walther-Mei{\ss}ner-Institut, Bayerische Akademie der Wissenschaften, 85748 Garching, Germany}
\affiliation{Technical University of Munich, TUM School of Natural Sciences, Physics Department, 85748 Garching, Germany}

\author{Manuel Müller}
\affiliation{Walther-Mei{\ss}ner-Institut, Bayerische Akademie der Wissenschaften, 85748 Garching, Germany}
\affiliation{Technical University of Munich, TUM School of Natural Sciences, Physics Department, 85748 Garching, Germany}

\author{Matthias Althammer}
\affiliation{Walther-Mei{\ss}ner-Institut, Bayerische Akademie der Wissenschaften, 85748 Garching, Germany}
\affiliation{Technical University of Munich, TUM School of Natural Sciences, Physics Department, 85748 Garching, Germany}

\author{Stephan Geprägs}
\affiliation{Walther-Mei{\ss}ner-Institut, Bayerische Akademie der Wissenschaften, 85748 Garching, Germany}

\author{Rudolf Gross}
\affiliation{Walther-Mei{\ss}ner-Institut, Bayerische Akademie der Wissenschaften, 85748 Garching, Germany}
\affiliation{Technical University of Munich, TUM School of Natural Sciences, Physics Department, 85748 Garching, Germany}
\affiliation{Munich Center for Quantum Science and Technology (MCQST), 80799 Munich, Germany}

\author{Friedemann Reinhard}
\email{friedemann.reinhard@uni-rostock.de}
\affiliation{University of Rostock, Institute for Physics, 18059 Rostock, Germany}
\affiliation{University of Rostock, Department of Life, Light and Matter, 18059 Rostock, Germany}
\affiliation{Munich Center for Quantum Science and Technology (MCQST), 80799 Munich, Germany}

\date{\today}

\begin{abstract}
Planar scanning probe microscopy is a recently emerging alternative approach to tip-based scanning probe imaging. It can scan an extended planar sensor, such as a polished bulk diamond doped with magnetic-field-sensitive nitrogen-vacancy (NV) centers, in nanometer-scale proximity of a planar sample. So far, this technique has been limited to optical near-field microscopy, and has required nanofabrication of the sample of interest. Here we extend this technique to magnetometry using NV centers, and present a modification that removes the need for sample-side nanofabrication. We harness this new ability to perform a hitherto infeasible measurement - direct imaging of the three-dimensional vector magnetic field of magnetic vortices in a thin film magnetic heterostructure, based on repeated scanning with NV centers with different orientations within the same scanning probe. Our result opens the door to quantum sensing using multiple qubits within the same scanning probe, a prerequisite for the use of entanglement-enhanced and massively parallel schemes.
\end{abstract}

\maketitle

\section{Introduction}
Imaging of magnetic fields with nanoscale resolution is an important challenge for various areas of science, including the study of nano-magnetism~\cite{tschudin_imaging_2024}, material science~\cite{palm_observation_2024}, and the design of hard-drive write heads~\cite{jakobi_measuring_2017}. 
Among the many techniques for magnetic field imaging~\cite{christensen_2024_2024} scanning-probe imaging using nitrogen-vacancy (NV) centers  has become a highly attractive method because it is quantitative, offers nanoscale spatial resolution, high sensitivity, operates over a wide range of experimental conditions (cryogenic to room temperature) and is more easily accessible than synchrotron-based methods like x-ray ptychography~\cite{donnelly_highresolution_2016}.
The most established approach to NV magnetic field imaging relies on nanofabricated diamond atomic force microscopy (AFM) tips, hosting a NV center at their apex~\cite{balasubramanian_nanoscale_2008, maletinsky_robust_2012}.
These tips are, however, difficult to fabricate and operate with. Furthermore, nanofabrication degrades the spin properties of NV centers, so that the technique has been mostly limited to sensing of strong DC magnetic fields.
Also, tip-based scanning probe experiments so far cannot leverage a key strength of NV centers - their ability to perform vector magnetometry, i.e. to measure all three components of the magnetic field vector. A tip capable of nanoscale imaging can only host a single NV center while the most established technique for vector magnetometry uses NV centers of different crystallographic orientations, exploiting that every orientation is predominantly sensitive to the magnetic field component along this particular axis~\cite{steinert_high_2010, maertz_vector_2010}. 
It should be noted that several schemes have been devised to enable vector magnetometry using NV centers of a single orientation. In an indirect approach, the vacuum Maxwell equations $\vec \nabla \cdot \vec B=0$ and $\vec \nabla \times \vec B = 0$ can be harnessed to compute the full vector field from a single-component magnetic field image~\cite{lima_obtaining_2009}. It has, however, been found that a direct measurement of all three components is superior for the reconstruction of source quantities like charge current or magnetization~\cite{broadway_improved_2020}. Also, using a single NV center with a fixed axis creates blind spots in large orthogonal fields, where the magnetic field sensitivity decreases drastically~\cite{rondin_magnetometry_2014}. 
Direct vector measurements of near-DC magnetic fields using a single NV center have also been demonstrated, based on modulating an external bias field~\cite{zheng_microwavefree_2020, tsukamoto_vector_2021} or conversion of magnetic DC to AC fields by a nearby nuclear spin~\cite{liu_nanoscale_2019}.
These schemes, however, are technically complex and have not yet been combined with scanning probe microscopy. As a result, direct NV vector magnetometry at the nanoscale has so far remained elusive.\\
Here, we demonstrate nanoscale scanning probe vector magnetometry using NV centers. This advance is based on planar scanning probe microscopy~\cite{ernst_planar_2019, fringes_nanofluidic_2018, salihoglu_nearfield_2020, silva_scanning_1994} - a recently introduced simplified scanning probe technique which obviates the need for nanofabricated tips and instead scans an extended planar sensor relative to a planar sample.
\section{Planar scanning probe microscopy without sample-side microfabrication}
\begin{figure}
    \centering
    \includegraphics[scale=1]{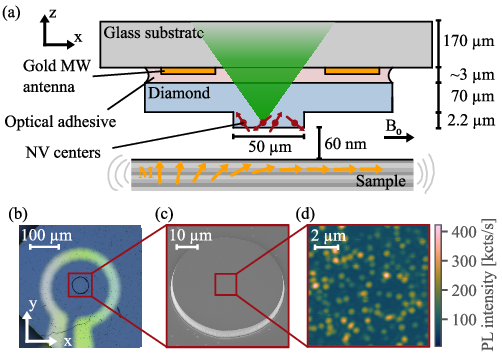}
    \caption{\label{fig:probe}(a) Schematic cross section of the planar scanning probe and sample. A laterally extended diamond pedestal with NV centers at the surface is scanned in close proximity to the magnetic multilayer sample. The optical excitation and readout is performed from the top, through the whole planar probe structure, which includes a microwave (MW) antenna and glass substrate for mounting the diamond. A small bias field $B_0$ was applied by external permanent magnets. (b) Optical microscope image of the planar scanning probe. The MW loop antenna can be seen underneath the thin diamond substrate. (c) Scanning electron microscope image of the etched diamond pedestal. (d) Confocal photoluminescence (PL) image of the NV centers inside the probe.}
\end{figure}

\begin{figure*}
    \centering
    \includegraphics[scale=1]{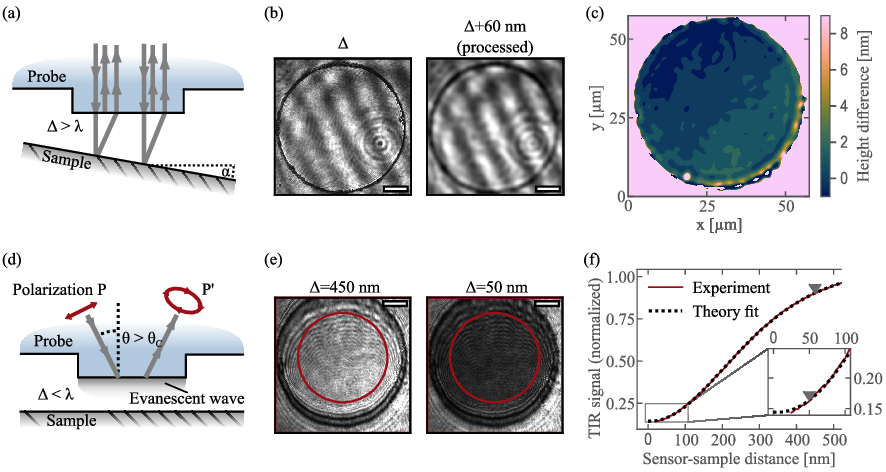}
	\caption{\label{fig:tiltdistance}(a) Tilt measurement technique using wide-field illumination of the pillar at a large stand-off distance $\Delta>\lambda$. The beams reflected at the diamond and sample surface interfere to create an interference pattern that encodes the tilt angle $\alpha$. (b) Interference image at a tilt angle before (left) and after (right) image processing. Images taken at $\Delta\approx 1\,\mathrm{\upmu m}$. The scale bars correspond to $10\,\mathrm{\upmu m}$. (c) Map of the height difference over the pillar surface after final alignment. (d) Distance measurement technique. Illumination is adjusted to an incidence angle larger than the critical angle of the diamond-air interface $\theta_\mathrm{c}$. The distance can be determined by measuring changes in the polarization of the reflected light. (e) Images of the pillar surface for different distances between sample and probe. For the distance measurement the reflected intensity is averaged over the circular region of interest. The scale bars correspond to $10\,\mathrm{\upmu m}$. (f) The stand-off distance is determined by fitting the total internal reflection (TIR) intensity while approaching the sample. The gray triangles indicate the distances, where the images in (e) where taken.}
\end{figure*}

Approaching a planar probe into nanometer-scale proximity of a planar sample is possible, if the sensor and the sample are aligned sufficiently parallel to each other. Ensuring such an alignment is possible by interferometric microscopy. Scanning probe imaging using NV centers has already been demonstrated using this technique~\cite{ernst_planar_2019}, but has so far been limited to optical near-field microscopy. Also, in past experiments the sample was placed on a $10\,\mathrm{\upmu m}$ sized microfabricated pillar, so that the technique could not be applied to arbitrary samples without additional sample-side fabrication. Here, we overcome these limitations, enabling planar scanning probe magnetometry on arbitrary samples. Our setup is presented in Fig.~\ref{fig:probe}(a). We approach and scan an extended diamond membrane within nanometer-scale proximity of a planar sample.
\\
A pedestal of $50\,\mathrm{\upmu m}$ diameter and $2.2\,\mathrm{\upmu m}$ height is etched into the diamond membrane by reactive ion etching to relax the demands on alignment accuracy and cleanliness of the sample. In return, this obviates sample-side microfabrication and enables work on arbitrary samples without additional preparation. NV centers are shallowly implanted at a nominal depth of $8\,\mathrm{nm}$ into the surface of the planar probe. Crucially, the pedestal can hold hundreds of NV centers, as can be seen in the photoluminescence (PL) image of the pedestal in Fig.~\ref{fig:probe}(d). All of these NV centers can be addressed individually by high-NA optics looking from the top of Fig.~\ref{fig:probe}(a), so that a sample can be repeatedly scanned with NV centers of all four different orientations, enabling vector magnetometry in the most direct way possible. 
While fabrication of the pedestal still is a nanofabrication step, it is considerably simpler than the process required to create diamond tips and integrate them with an AFM cantilever. Also, NV centers in the pedestal can be several microns away from any edge influenced by the fabrication and thus retain excellent charge stability and spin properties.\\
In order to achieve a small stand-off distance with this planar probe, parallel alignment between it and the sample surface is crucial.\\
The relative sensor-sample angle is measured by illuminating the probe with a collimated beam of light from above as shown in Fig.~\ref{fig:tiltdistance}(a), such that there will be reflections at the probe and sample surface, that interfere to form fringes similar to Newton's rings. Large sample-probe angles can be measured quantitatively from just a single image  as depicted in Fig.~\ref{fig:tiltdistance}(b) and then corrected by tilting the sample with a piezo tip-tilt positioner. For smaller angles during the fine tilt adjustment the size of the interference fringes becomes larger than the pedestal diameter, making tilt measurement from a single image infeasible. Instead, we gradually move the probe upwards to increase the sensor-sample distance $\Delta$ while acquiring images at each height. For each point on the pedestal we record the periodic intensity variations caused by interference. Comparing the intensity curves of different points on the pedestal will reveal a small offset between the curves caused by a difference in sensor-sample height at these points, which is shown in Fig.~\ref{fig:tiltdistance}(c).
\\
Due to the wide-field illumination from above, reflection at the sidewall of the pedestal and other interfaces along the optical path cause additional interference patterns in the image, which impair the tilt measurement. Using suitable spatial low-pass filters for each image as well as filtering the intensity variations along $\Delta$ for each pixel reduces the influence of these parasitic inferences and other noise like laser power fluctuations. After the final alignment the maximum height difference was only $2\,\mathrm{nm}$ over the $50\,\mathrm{\upmu m}$ pedestal surface, which even improves upon the results in a previous work with this technique~\cite{ernst_planar_2019}. This result also shows that, surprisingly, the roughness and curvature of a polished diamond membrane can be controlled to this nanoscale level, even when mounted in a setting that generates additional bending forces, e.g. from the oil immersion objective pushing from above in Fig.~\ref{fig:probe}(a).\\
While interferometry can be used for measuring changes in the sample-probe distance, it does not provide information about the absolute distance. Therefore we use an all-optical, non-contact distance measurement technique based on Brewster angle microscopy with sub-nanometer resolution for distance control in the last $\approx 100$ nm of the final approach. Oblique illumination with linearly polarized light is used to create total internal reflection at the diamond-air interface. Upon reflection the polarization is rotated by $90^\circ$. When approaching the sample to a distance smaller than the light's wavelength the outgoing polarization starts to become elliptic, which can be understood as interaction of the evanescent wave with the sample. The change in polarization can be translated to a change in the reflected intensity in a cross polarization setup. Upon approaching the probe to the sample this change in intensity is recorded and fitted to an analytic expression derived with a transfer-matrix approach~\cite{ernst_planar_2019}. The sensor-sample distance is then extracted as a fit parameter.\\
During a scanning probe measurement this method is also used as a feedback signal for a software-based control loop to counteract drifts of the stand-off distance every $\approx 3\,\mathrm{s}$.
\section{Single NV center magnetic imaging}
\begin{figure*}
    \centering
    \includegraphics[scale=1]{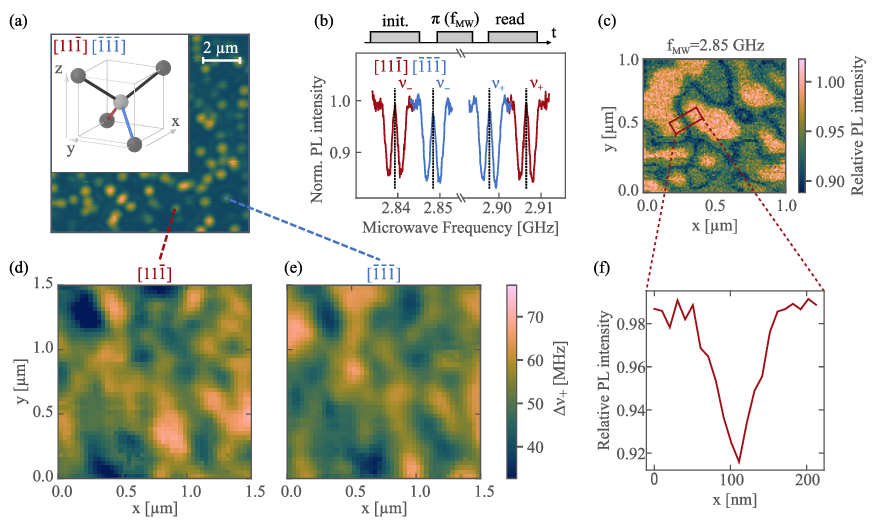 }
    \caption{\label{fig:isoBmap}(a) Four crystallographic orientations of the NV center. The PL image shows two NV centers of different orientation, which were used for the magnetic field imaging measurements. (b) ODMR pulse sequence and spectrum for two NV centers. Due to the different angle towards the bias field vector the frequency splitting $\Delta\nu$ is different. The double dip for each transition frequency is caused by the hyper-fine splitting from implantation of $^{15}\mathrm{N}$ ions. (c) Magnetic isoline image taken at $60\,\mathrm{nm}$ stand-off distance with a microwave frequency of $2.85\,\mathrm{GHz}$, corresponding to a magnetic field of $0.71\,\mathrm{mT}$. The NV center was oriented along $\left[\,1,1,\overline{1}\,\right]$. (d) and (e) Map of the shift of $\nu_+$ when scanning different NV centers over the sample at a distance of $80\,\mathrm{nm}$. (f) Line profile over the outlined area in (c).}
\end{figure*}

We performed magnetometry on a magnetic multilayer structure of Pt, CoFe, and Ir thin films with easy-plane anisotropy. Due to the interfacial Dzyaloshinskii-Moriya interaction (iDMI) these multilayers can host a skyrmion phase at room temperature (see Ref.~\cite{flacke_robust_2021} for further details on the sample fabrication and magnetic properties). Their non-uniform magnetization makes them a prime candidate for vector magnetic field imaging.\\
For the magnetometry a pulsed optically detected magnetic resonance (ODMR) measurement with the sequence shown in Fig.~\ref{fig:isoBmap}(b) was employed. With a bias field of known orientation and ODMR measurements of the hundreds of NV centers in the planar probe it was possible to determine their crystallographic direction. Additionally the $T_2$-time of all NV centers can be measured beforehand to select suitable ones. This marks another simplification over tip-based schemes where a whole new probe needs to be fabricated in case of poor spin properties or wrong NV center orientation.
\\
Firstly a B-field isoline measurement~\cite{balasubramanian_nanoscale_2008} was performed by scanning the NV center over the sample at a stand-off distance of $60\,\mathrm{nm}$ while repeatedly running a pulsed ODMR measurement at a fixed frequency. Areas of the sample where the Zeeman splitting due to the magnetic stray field coincides with the driving MW frequency show reduced PL intensity as illustrated in Fig.~\ref{fig:isoBmap}(b). This is a standard measurement technique in tip based scanning probe experiments, where only a single magnetic field component along the NV axis (in this case $\left[\,1,1,\overline{1}\,\right]$) is measured and full information of the magnetic field is sacrificed in favor of a fast image acquisition time.\\
A complex, maze-like magnetic field isoline can be seen, where domains down to $\approx 100\,\mathrm{nm}$ could be resolved. Importantly the isolines' width does not represent the spatial resolution of our technique. The width is given by the ODMR linewidth and the gradient of the magnetic field. The line profile in Fig.~\ref{fig:isoBmap}(f) through the isoline shows a full width at half maximum (FWHM) of $57\,\mathrm{nm}$, which, can be converted to a magnetic field gradient of $3.9\,\frac{\mathrm{mT}}{\mathrm{\upmu m}}$ with the ODMR linewidth of $6.2\,\mathrm{MHz}$ FWHM.\\
While the quantification of the spatial resolution would require a priori knowledge of the underlying magnetization, it has been shown that the resolution is approximately equal to the distance between NV center and the sample~\cite{hingant_measuring_2015}, which was $68\,\mathrm{nm}$ for this measurement (sum of stand-off distance $\Delta$ and implantation depth).
\section{Vector magnetic field imaging}
\begin{figure}
    \centering
	\includegraphics[scale=1]{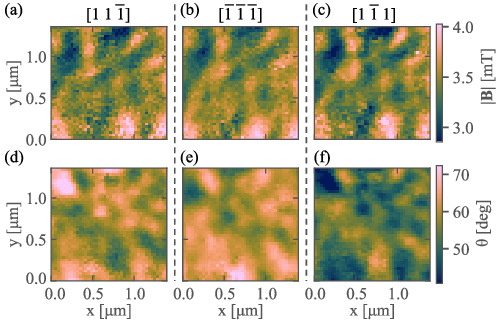}
	\caption{\label{fig:3Draw}(a)-(c) Maps of the magnetic field strength and (d)-(f) angle between magnetic field vector and NV center axis for different NV centers with their respective crystallographic axis shown above.}
\end{figure}

\begin{figure}[h]
    \centering
	\includegraphics[scale=1]{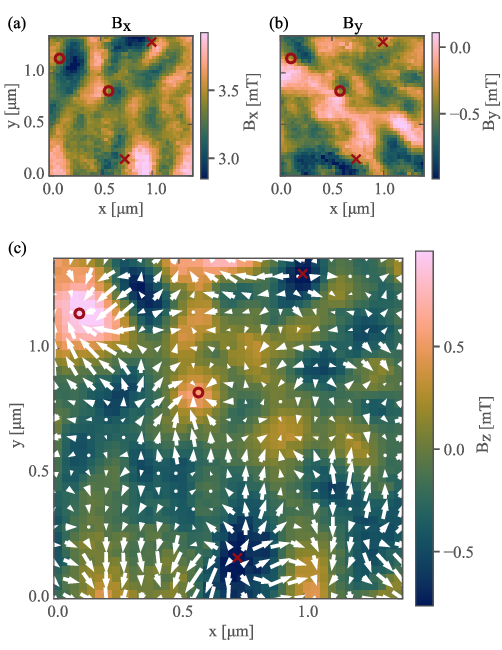}
	\caption{\label{fig:3Dres}(a) and (b) In-plane components of the reconstructed magnetic field vector. The center of magnetic vortices with positive (negative) z-component is indicated by red circles (crosses). (c) Out-of-plane component of the magnetic field with the in-plane field overlaid as white arrows. The bias field has been subtracted from the in-plane components to show the variation of magnetic field caused by the sample magnetization.}
\end{figure}

Finally we present the vector magnetic field imaging technique, which works by scanning an area with multiple NV centers with different, previously determined directions, thus measuring projections of the magnetic field onto the corresponding axes, which can then be combined into a 3D vector.
For each NV center we conduct a pulsed ODMR measurement at each pixel, during which both $\nu_-$ and $\nu_+$ are measured. The resulting images of $\nu_+$ for two different NV centers at a stand-off distance of $80\,\mathrm{nm}$ are shown in Fig.~\ref{fig:isoBmap}(d) and (e). The frequencies $\nu_-$ and $\nu_+$ are unambiguously defined by the magnitude of the magnetic field vector $\left|\mathbf{B}\right|$ and the angle $\theta$ between the magnetic field vector and the NV center axis. The energies corresponding to the two transitions are given by the eigenvalues of the Hamiltonian
\begin{equation}
    H = hD\mathbf{S_z}^2+h\gamma\left|\mathbf{B}\right|\cos{\theta}\,\mathbf{S_z}+h\gamma\left|\mathbf{B}\right|\sin{\theta}\,\mathbf{S_x}
    \label{eq:Hamiltonian}
\end{equation}
where $D=2.87\,\mathrm{GHz}$ is the zero-field splitting, $\mathbf{S_z}$ and $\mathbf{S_x}$ are two Pauli matrices, $h$ is the Plank constant and $\gamma=28\,\frac{\mathrm{MHz}}{\mathrm{mT}}$ the gyromagnetic ratio of the NV center.
With this equation a nonlinear least-squares fit of the measured $\nu_-$ and $\nu_+$ can be used to find the corresponding magnetic field strength $\left|\mathbf{B}\right|$ and angle $\theta$. This resonance frequency measurement and fitting routine is then repeated over the same area of the sample, but with three different NV centers with different, previously determined orientations, resulting in images of $\left|\mathbf{B}\right|$ (Fig.~\ref{fig:3Draw} (a)-(c)) and $\theta$ (Fig.~\ref{fig:3Draw}(d)-(f)).
\\
Special care has to be taken that the same area is scanned with the three different NV centers that have a relative distance of few micrometers. We achieve this by a combination of two approaches. Firstly stochastic optical super-resolution imaging of the diamond probe is used to determine the position of NV centers with sub-diffraction-limited resolution. The sample is then moved so that every NV center is scanned over the same area.
Secondly, we extract the resulting images of $\left|\mathbf{B}\right|$ from all three orientations (Fig.~\ref{fig:3Draw}(a)-(c)) and use them to fit and correct a small remaining lateral shift relative to each other. This shift was determined by performing image registration, where we minimize the L1 norm of the difference between consecutive images by performing sub-pixel shifts in $x$ and $y$ direction. The images of the measured resonance frequencies (examples for $\nu_+$ in Fig.~\ref{fig:isoBmap}(d) and (e)) were then shifted to compensate for this remaining offset.\\
The reconstruction of the magnetic field vector can be done via the simple geometric consideration that for each NV center the projection of the magnetic field vector $\left(B_x, B_y, B_z\right)$ onto the NV center axis is given by the product of magnetic field strength and angle $\theta$, e.g. for the $\left[\,1, 1, \overline{1}\,\right]$ direction:
\begin{equation}
\frac{1}{\sqrt{3}}
\left(\begin{array}{c}
    1 \\
    1 \\
    -1 
    \end{array} \right)
\cdot
\left(\begin{array}{c}
    B_x \\
    B_y \\
    B_z 
    \end{array} \right)
= \left|\mathbf{B}\right|
\cos{\theta_{\left[1,1,\overline{1}\right]}}
\label{eq:projection}
\end{equation}
Combining corresponding equations for the three different NV center directions gives a system of linear equations that theoretically can be solved directly to find the $\mathbf{B}$ vector.\\
However we find that this analytic approach does not work in the presence of noise in $\left|\mathbf{B}\right|$ and $\theta$.
Instead, we use nonlinear least-squares fitting to find the magnetic field vector that matches the measured resonance frequencies. The three cartesian components of this solution are shown in Fig.~\ref{fig:3Dres}.
\\
We verify the reconstructed magnetic field vector by calculating the resonance frequencies for the three different NV orientations using Eq.~\ref{eq:Hamiltonian}. A comparison to the measured values shows an excellent agreement with a normalized-root-mean-square deviation of $7.5\cdot10^{-4}$.\\
Analyzing the resulting magnetic field vector we find that the strongest magnetic field component is $B_x$, because it almost fully coincides with the applied bias field direction. However all three components show a non-trivial nanoscale magnetic field structure with $\approx 1\,\mathrm{mT}$ amplitude. For the bias field amplitude in this work previous magnetic force microscopy (MFM) measurements of this structure~\cite{flacke_robust_2021} have revealed a similar maze domain structure although with much smaller domain size.\\
Combining the components to a vector field plot in Fig.\,~\ref{fig:3Dres}(c) reveals several vortex structures in the magnetic field. These vortices have a size of $\approx 200\,\mathrm{nm}$ and exhibit a maximum or minimum in the out-of-plane magnetization combined with a change of direction of the in-plane components. While such a field is a general feature seen over a magnetic monopole, for example one end of an elongated in-plane magnetic dipole, this signature is also consistent with Néel-type magnetic skyrmions, which have been reported in this structure~\cite{flacke_robust_2021}.
\section{Conclusion}
In summary, we have successfully pushed planar scanning probe microscopy to full scanning NV center magnetometry with a diamond pedestal that removes the need for sample-side microfabrication. The resulting technique requires significantly less complex fabrication compared to diamond scanning tips. We demonstrated the viability of our technique by imaging the nanoscale magnetic field of a magnetic multilayer structure with magnetic vortices, demonstrating direct vector magnetic field imaging at the nanoscale.\\
This work also is, to the best of our knowledge and together with recent tip-based work~\cite{huxter_multiplexed_2024}, the first demonstration of NV magnetometry that actively employs multiple individually addressable sensing qubits in the same scanning probe, not merely an average signal of an NV ensemble~\cite{laraoui_imaging_2015, tetienne_scanning_2016, simon_directional_2021, liu_nanoscale_2023}. This ability to employ several centers for sensing in the same scanning probe opens the door to a wide range of powerful schemes. Magnetic-dipole-coupled NV pairs~\cite{dolde_roomtemperature_2013} could be employed for entanglement-enhanced imaging of magnetic field gradients. Correlation-based protocols between separated NV centers could be employed to image higher-order spatial correlations of magnetic fields~\cite{rovny_nanoscale_2022}, e.g. to measure and image the nature and lifetimes of the excitations involved in charge transport~\cite{zhang_nanoscale_2024}. Finally, simultaneous readout of all NV centers in the planar scanning probe by a wide-field camera would enable magnetometry with simultaneously $10\,\mathrm{nm}$-scale resolution and a $100\,\mathrm{\upmu m}$-sized field of view, enabling fast scanning of extended magnetic field maps with tens of megapixels of resolution, comparable to the best present (4K and 8K) displays.
\section*{Methods}
The starting material for the planar probe was a surface-polished ($\mathrm{Ra}\!<\!5\,\mathrm{nm}$), electronic grade bulk diamond with $[100]$ face orientation grown by chemical vapor deposition, which was purchased from Element Six. NV centers were created close to the diamond surface by implantation of nitrogen ions with a dose of $5\cdot10^9\,\mathrm{ions/cm^2}$ and an energy of $5\,\mathrm{keV}$, followed by annealing at $830\,^\circ\mathrm{C}$ for $4$ hours. The resulting areal NV center density of $1.0\,\mathrm{\upmu m}^{-2}$ was chosen, so that isolated, single NV centers could be resolved optically (Fig.~\ref{fig:probe}(d)).\\
The pedestal was then fabricated via the following process: A $200\,\mathrm{nm}$ silicon nitride (SiN) layer was deposited via inductively-coupled plasma-enhanced chemical vapor deposition (ICPECVD) (Sentech SI 500 D). In a next step spin-coating of $1.4\,\mathrm{\upmu m}$ optical resist (AZ5214E Merck Performance Materials GmbH) and laser lithography (Heidelberg Instruments DWL66+) was performed to structure the circular $\diameter\,50\,\mathrm{\upmu m}$ resist mask. Subsequently, the SiN layer was structured via a 60 second sulfur hexafluoride (SF$_6$) inductively coupled plasma (ICP) dry etching step (Sentech SI 500). Finally, the diamond pedestal was created via oxygen ICP etching at $30\,\mathrm{sccm}$ oxygen flow, base pressure of $1.3\,\mathrm{Pa}$, ICP power of $700\,\mathrm{W}$ and platen power of $45\,\mathrm{W}$. The total etch depth after 12 minutes process time was determined to be $2200\,\mathrm{nm}$.\\
Optical excitation and readout of the NV centers is performed through the planar probe by a NA 1.42 objective. Because of the short working distance of the microscope objective and optical aberrations induced by imaging through the diamond with a high refractive index, the sample was thinned down to a thickness of $70\,\mathrm{\upmu m}$ by polishing the backside (performed by Almax easyLab bv). An optical adhesive (Norland NOA 63) was then used to glue the thin diamond membrane to a $170\,\mathrm{\upmu m}$ thick glass wafer on which a microwave antenna was fabricated beforehand.\\
The microwave antenna was a lithographically defined gold loop with diameter of $\approx 230\,\mathrm{\upmu m}$ and film thickness of $200\,\mathrm{nm}$.\\
Piezo positioning stages were used to move the planar probe in three dimensions relative to the optical axis in order to perform PL imaging of the diamond probe. A second piezo stage with three translation axes and two tilt axes was used to adjust tip, tilt, distance and lateral position of the sample with respect to the probe.

\section*{Data availability statement}
The data cannot be made publicly available upon publication because they are not available in a format that is sufficiently accessible or reusable by other researchers. The data that support the findings of this study are available upon reasonable request from the authors.

\section*{Acknowledgement}
This work has been supported by the Deutsche Forschungsgemeinschaft (DFG, grants RE3606/1-2, RE3606/3-1 and excellence cluster MCQST EXC-2111-390814868, SFB 1477 “Light–Matter Interactions at Interfaces” (Project No. 441234705)) and the European Union (ASTERIQS, Grant Agreement No. 820394).

\section*{Author contributions}
P.W. and F.R. conceived the project. P.K. and C.G. fabricated the diamond pedestal. L.F. fabricated and characterized the magnetic multilayer sample under supervision from S.G. and M.A. P.W. fabricated the planar scanning probe device, built the setup and performed the experiment under supervision from F.R. P.W., L.F., M.M., M.A., S.G., R.G. and F.R. analyzed the data. P.W. and F.R. wrote the paper. All authors read and commented on the final manuscript.

\section*{Competing interests}
The authors declare no competing financial or non-financial interests.

\section*{References}
\bibliographystyle{iopart-num.bst}

\providecommand{\newblock}{}

\end{document}